Publication date: 25 September 2022

European Union, Horizon 2020, Grant Agreement number: 857470 — NOMATEN — H2020-WIDESPREAD-2018-2020

This article appeared in Surface and Coatings Technology, Volume 446, 128802 (2022) and may be found at https://doi.org/10.1016/j.surfcoat.2022.128802

# Multi-component low and high entropy metallic coatings synthesized by pulsed magnetron sputtering

G. W. Strzelecki[1], K. Nowakowska-Langier[1], K. Mulewska[1,2], M. Zieliński[2], A. Kosińska[1,2], S. Okrasa[1], M. Wilczopolska[1,2], R. Chodun[3], B. Wicher[3,4], R. Mirowski[1], K. Zdunek[3]

[1] National Centre for Nuclear Research (NCBJ), Material Physics Department, Andrzeja Sołtana 7, 05-400, Otwock, Poland
[2] National Centre for Nuclear Research, NOMATEN Centre of Excellence, A. Sołtana 7, 05-400 Otwock-Świerk, Poland
[3] Faculty of Materials Science and Engineering, Warsaw University of Technology, Wołoska, 141 02-507 Warsaw, Poland
[4] Thin Film Physics Division, Department of Physics (IFM), Linkoping University, SE-581 83 Linkoping, Sweden

Corresponding author: G. W. Strzelecki – grzegorz.strzelecki@ncbj.gov.pl

Abstract

This paper presents the findings of the synthesis of multicomponent (Al, W, Ni, Ti, Nb) alloy coatings from mosaic targets. For the study, a pulsed magnetron sputtering method was employed under different plasma generation conditions: modulation frequency (10 Hz and 1000 Hz), and power (600 W and 1000 W). The processes achieved two types of alloy coatings, high entropy and classical alloys. After the deposition processes, scanning electron microscopy, X-ray diffraction, and energy-dispersive X-ray spectroscopy techniques were employed to find the morphology, thickness, and chemical and phase compositions of the coatings. Nanohardness and its related parameters, namely $H^3/E_r^2$, $H/E$, and $1/E_r^2 H$ ratios, were measured. An annealing treatment was performed to estimate the stability range for the selected coatings. The results indicated the formation of as-deposited coatings exhibiting an amorphous structure as a single-phase solid solution. The process parameters had an influence on the resulting morphology—a dense and homogenous as well as a columnar morphology, was obtained. The study compared the properties of high-entropy alloy (HEA) coatings and classical alloy coatings concerning their structure and chemical and phase composition. It was found that the change of frequency modulation and the post-annealing process contributed to the increase in the hardness of the material in the case of HEA coatings.

High entropy alloys

1. Introduction

There is a constant demand for alloys with specific or tailored properties—such as mechanical resilience [1,2], creep resistance [3], or corrosion resistance [4]. Currently, research is being conducted on more complex materials, that is, multiprincipal alloys with simple (single or dual) phase composition. There have been many studies conducted to create new metallic materials using the synthesis method by combining two or more elements. These studies have reported nonequilibrious results, such as rapid cooling, mechanical alloying, and vapor deposition. Some studies have also attempted to extend the ranges of occurrence of single-phase nonequilibrium areas [5–7] in the synthesis of two immiscible elements or the synthesis of complex metallic coatings (CoCrAl, FeCoNiAlCuTi) with impulse plasma from multicomponent sources [8,9]. Multiprincipal (equimolar) alloys, which were introduced in 1996 [10], were called high-entropy alloys (HEAs) [11,12]. Then defined as an equimolar alloy of 5 or more elements and now accepted as an alloy of 5 or more elements ranging from $5\%_{at}$ to $35\%_{at}$ are proving to be the next step in material design. Although HEAs are a relatively new group of materials, they have certain advantages over classical alloys. The cocktail effect [13,14] creates different alloys and derived materials, such as oxides [15], nitrides [16–18], and carbides [19]. Therefore, the HEAs can exhibit properties such as corrosion resistance [20], catalytic activity [21], and high hardness [22]. As a result of increased difficulty of diffusion within the material and an increased internal entropy [13], these alloys also have heightened phase and structural thermal stability [23–25]. This helps in creating alloys that are designed to work at elevated temperatures. Since alloyed targets, mosaics (multitarget) [26], and multimagnetron [27] setups are well known, HEAs can be easily introduced in magnetron sputtered coatings [28]. The synthesis technique also creates multicomponent coatings in several chemical compositions. However, using upscaled industrial multitarget systems pose a challenge in retaining the desired chemical composition. Furthermore, the use of multiple magnetrons in vacuum systems is associated with additional challenges; since more elements are needed to create a coating, the bigger the vacuum chamber, the higher its cost. The economic impact of multielement coating systems can be lessened by using mosaics. However, prior calculations are required for using multimagnetron systems and mosaic targets to ensure the correct chemical compositions of the synthesized coatings.

In this study, an additional parameter of synthesis [29,30]—modulation frequency ($f_{mod}$)—is used for pulsed magnetron sputtering (PMS). The modulation frequency, schematically presented in Figure 2., changes how energy is fed into the sputtering system, and thus how plasma is generated during sputtering processes. Additionally, the figure shows a zoomed-in fragment of the internal 100 kHz excitation frequency. Because of this, there is better control of coating growth [31], and therefore, a wide range of possible structures and properties, such as electrical resistivity [32,33], optical properties and hardness [34], structure [33,35], texture [35,36], and phase composition [37] can be achieved.

This paper presents the results of two groups of coatings sputtered using two different mosaic targets, with different chemical compositions of the synthesized coatings. The coatings were synthesized under various conditions of plasma generation, i.e., power and frequency modulation ($f_{mod}$). The study examined the morphology and phase compositions, and the structure's influence on their properties by nanoindentation behavior. In addition, attempts were made to obtain information about the possibility of creating single-phase alloys by applying a specific feature of the PMS method, which can control the frequency of pulse plasma generation.

2. Experimental

2.1 Target design

The two mosaic targets manufactured and used in the study are presented in Figure 1. The relative surface areas of each element and modulation frequency were used to control the chemical composition of the synthesized coatings. However, synthesis has a few disadvantages concerning the consumption of the target and the changes in the relative share of the sputtered surface [38]. Therefore, calculations presented by Dolique et al. [39] were used in the study. Because of the degradation of the target, the study of each target was kept to a minimum. The results of the initial calculation are shown in Table 1.

Since the components of HEAs determine the properties of the resulting alloys and the synthesis conditions, it is important to select the correct components. The component metals of the layers were chosen to synthesize the lightweight HEA materials with broad application prospects, such as in the aviation industry [40]. The common structural alloys in such applications are based on one principal element such as Al, Ti, Fe, or Ni, and other minor elements are added as alloys to modify the microstructure and the properties. The alloying additives used in this study were selected based on their properties of improving the hardness, which is exhibited by titanium, as shown by Zhou et al. [41]. Previous studies [42] have shown that the addition of niobium also increases the hardness. Additionally, metals Nb and W, which are commonly used in refractory alloys, were used to increase the thermal stability [43] of the synthesized samples.

The mosaics created had an aluminum (Al) base and inserts of tungsten, titanium, niobium, and nickel. Since Al has the highest thermal conductivity coefficient value, is of low cost, and allows easy machining, it was used to aid in the overall cooling of the sputtering system. The metallic inserts were evenly spaced within the sputtering racetrack of the target. Two different designs were employed, as seen in Figure 1(a) and (b). The first (a) design contained a relatively low concentration (35%–40%) of insert metals in resultant coatings, while the second (b) design contained about 70% of insert metals in the synthesized coatings. The addition of a ferromagnetic metal (Ni) in the assembly necessitated the changes in the inserts' size, shape, and location to minimize the distortions of the local magnetic field.

The mixing entropy ($\Delta S_{mix}$) [44] was calculated for each instance according to equation (1) to verify the performance of the target design and ensure that sputtered coatings were of the desired type, as shown in Table 1. Equation (1): R is the ideal gas constant and $n_k$ is the mole fraction of the $k$th element. Since previous findings show several ways of defining HEAs as a function of their $\Delta S_{mix}$, this study used a noncontroversial delineation, with $\Delta S_{mix} \geq 1.5R$ indicating HEAs and $\Delta S_{mix} \leq 1R$ indicating low-entropy alloys or classical alloys [45,46].

$$\Delta S_{mix} = -R \sum_{i=1}^{j} n_k \ln(n_k) \quad (1)$$

The valence electron concentration (VEC) has been shown [47] to be a useful parameter for determining whether the fcc or bcc phase will be stable in the synthesized material and as such has been calculated for high entropy films as per equation (2) and is shown in Tab. 1. Equation (2): $VEC_k$ is the valence electron concentration of the $k$th element and $n_k$ is the mole fraction of the $k$th element.

$$VEC = \sum_{n=1}^{k} n_k (VEC)_k \quad (2)$$

Atomic size difference (Eq.3) and enthalpy of mixing (Eq.4) proposed by Zhang et al. [48] as an additional set of parameters used in high entropy alloy design.

$$\delta\% = 100\% \sqrt{\sum_{i=1}^{k} c_i (1 - \frac{r_i}{\sum_{j=1}^{n} c_j r_j})^2} \quad (3)$$

Where $r_i$ is the atomic radius of $i$th element and is the average radius.

$$\Delta H_{mix} = \sum_{i=1, i \neq j}^{n} 4 \Delta H_{ij}^{mix} c_i c_j \quad (4)$$

Where is the enthalpy of mixing of the binary alloy between the *i*th and *j*th elements at an equiatomic composition.

Electronegativity difference of multi-component alloy system is given by equation (4) as defined by Fang et al [49].

$$\Delta \chi = \sqrt{\sum_{i=1}^{n} c_i (\chi_i - \sum_{j=1}^{n} c_j \chi_j)^2} \quad (5)$$

where χ$_i$ is the electronegativity of the *i*th element, and is the average electronegativity.

Table 1. contains every value used in equations 2 – 5.

| Element | Atom radius [pm] [50] | $\Delta H_{ij}^{mix}$ [51] | | | | VEC [47] | Pauling electronegativity χ [47] |
|---|---|---|---|---|---|---|---|
| Al | 143.17 | Al-W | -13.80 | W-Ni | -3.13 | 3 | 1.61 |
| W | 136.7 | Al-Nb | -29.82 | W-Ti | -5.64 | 6 | 2.36 |
| Nb | 142.9 | Al-Ni | -32.55 | Nb-Ni | -29.51 | 5 | 1.6 |
| Ni | 124.59 | Al-Ti | -40.48 | Nb-Ti | -1.98 | 10 | 1.91 |
| Ti | 146.15 | W-Nb | -8.25 | Ti-Ni | -34.09 | 4 | 1.54 |

Table 1. Relevant thermodynamic and physiochemical properties of elements used in synthesis.

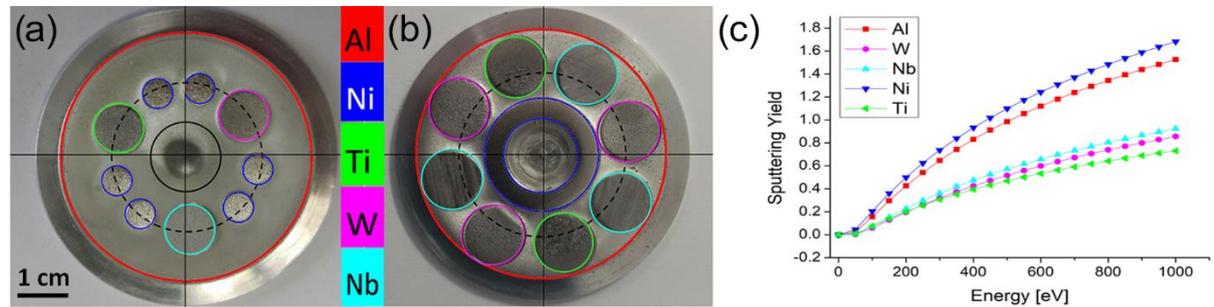

Fig. 1. – Pictures of used targets: classical (a) – designated as Mosaic 1, and HEA (b) – designated as Mosaic 2; (c) shows sputtering yield curves for selected metals.

| Target | Al | Ti | Ni | Nb | W | $\Delta S_{mix}$ [R] | $\Delta H_{mix}$ [kJ/mol] | VEC | δ [%] | $\Delta \chi$ |
|---|---|---|---|---|---|---|---|---|---|---|
| M1 | 70.08 | 6.04 | 16.37 | 3.59 | 3.92 | 0.96 | -28.51 | 4.40 | 4.99 | 0.18 |
| M2 | 28.16 | 14.44 | 21.82 | 16.10 | 19.48 | 1.58 | -33.48 | 5.58 | 5.63 | 0.30 |

Table 2. Calculated values of expected chemical composition and corresponding $\Delta S_{mix}$ –mixing entropy, $\Delta H_{mix}$ – mixing enthalpy, valence electron concentration, δ [%] – atomic size difference and $\Delta \chi$ – electronegativity difference.

Based on the results from equations (1)–(5), the theoretical morphology of the coatings should be clearly delineated between low ($\Delta S_{mix}$ < 1R) and high ($\Delta S_{mix}$ > 1.5R) entropy alloys. According to the VEC values, it has (<6.87) a bcc structure. However, as summarized by Zhang, very low values, close to or above 5%, of $\Delta H_{mix}$ and δ fall outside (–15 kJ/mol < $\Delta H_{mix}$ < –5 kJ/mol; 0 < δ < 5) of the single-phase solid solution formation region [52]. In a study conducted by Guo [53], it was observed that the chosen alloys will be metallic glasses as –49 kJ/mol ≤ $\Delta H_{mix}$ ≤ –5.5 kJ/mol, even though δ is significantly below

9. This is a distinct possibility since the condensing vapors created during sputtering undergo rapid cooling upon reaching the substrate.

2.2 Sputtering

Using a PMS system, multicomponent coatings were obtained on silicon Si (100) (Lukasiewicz Research Network—Institute of Electronic Materials Technology, Warsaw, Poland) wafers. Before sputtering, the silicon substrate was ultrasonically cleaned (Codyson 0.6l 50W CD3800, Shenzhen Codyson Electrical Co., Guangdong, China) in acetone (Linegal Chemicals, Warsaw, Poland), and was later installed on a grounded stage in the vacuum chamber.

This study employed the PMS technique [54,55]. During sputtering, a mix of high-purity (5N) argon with 5%$_{at}$ hydrogen was used. To maintain the high purity of the resulting coatings, hydrogen was added to chemically bond with the residual oxygen. The vacuum chamber was evacuated before each process to a base pressure of 3E-4Pa, and is further flooded with the Ar+H2 mix to purge the gas feeds and the chamber. The pressure was set to 5E-1Pa during sputtering. The coatings were deposited onto parallelly aligned Si (100) substrates placed on a grounded metallic sample holder, kept 100 mm from the sputtered target. The WMK50 circular magnetron (IMT, Wroclaw University of Technology, Poland) was powered by a Dora Power System (DPS) pulsed power supply. A DPS power supply allows adjusting the modulation frequency of plasma generation. The characteristic dependencies that accompany the PMS synthesis processes demonstrate the specificity of this process [36]. Figure 2. provides a schematic view of the plasma-generating signal for 10 Hz and 1000 Hz modulation frequency settings. No external sources were used to heat the substrates. The substrate temperature did not exceeded 100°C during the synthesis.

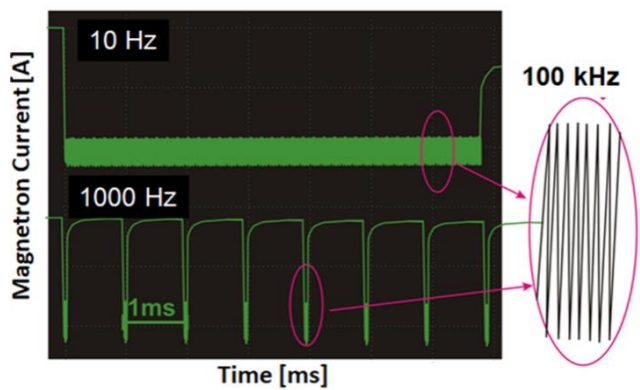

Fig. 2. Schematic view of the voltage characteristics that are typical of the PMS process.

Provided in Table 3. are synthesis parameters used during magnetron sputtering synthesis and clarification of the naming convention.

| Target | Power [W] | Sputtering time [min] | Modulation frequency [Hz] | Ar+H$_2$ pressure [Pa] | Names of as-deposited samples | Names of annealed samples | Annealing temperatures [°C] |
|---|---|---|---|---|---|---|---|
| Mosaic 1 | 1000 | 10 | 10 | 0.5 | M1.S1 | M1.S1A | 200 |
| Mosaic 1 | 1000 | 10 | 1000 | 0.5 | M1.S2 | M1.S2A | 400 |
| Mosaic 2 | 600 | 10 | 10 | 0.5 | M2.S1 | M2.S1A | 200 |
| Mosaic 2 | 600 | 10 | 1000 | 0.5 | M2.S2 | M2.S2A | 400 |
| Mosaic 2 | 1000 | 10 | 10 | 0.5 | M2.S3 | M2.S3A | 600 |

| Mosaic 2 | 1000 | 10 | 1000 | 0.5 | M2.S4 | M2.S4A | 800 |

Table 3. Synthesis parameters and annealing temperature for all coatings studied in this research paper and sample naming convention.

2.3 Sample preparation

The synthesized samples were divided into two research sets: as-deposited reference samples and annealed samples, as seen in Table 3. The XRD results of annealed M1.S1 and M2.S1 were used to determine the M1.Sx and M2.Sx annealing temperatures, respectively. The description of XRD annealing is presented in this paper (2.4 Sample characterization). The annealing processes were carried out in a furnace, under a constant flow of argon. The annealing temperatures were as follows: 200°C and 400°C for M1.Sx samples, and 200°C, 400°C, 600°C, and 800°C for M2.Sx samples.

2.4 Sample characterization

Scanning electron microscopy (SEM) images of the cross-sections of the obtained coatings were collected with low-energy (3 keV) electrons using a Helios 5 UX microscope (Thermo Fisher Scientific), equipped with a through-lens detector), which is designated for high-resolution imaging in the immersion mode. The EDX (Energy Dispersive X-ray) analysis was carried out on the Scanning Electron Microscopy (SEM) provided by Zeiss EVO MA 10. The device was equipped with a secondary electron Bruker XFlash Detector 5010. Using cross-sectional views, the thickness of coatings and growth rate were estimated, and the results are shown in Table 5.

Diffraction patterns (DPs) acquisition for M1.Sx and M2.Sx samples were performed using a divergent X-ray beam and the Bragg-Brentano geometry at the Bruker D8 Advance diffractometer with θ/θ goniometer of 280 mm radius. An offset of 2.5° was introduced for HEA samples to hide the Si peak. The intensity of the Si (4 0 0) reflection was very weak with the addition of the offset, which also prevented the saturation of the detector. The X-ray tube had a Cu anode powered by 40 kV and 40 mA. Data were collected by LYNXEYE XE-T detector in the high-resolution 1D mode without the Ni-filter, and its matrix (192 strips, each 0.075 mm wide) covered 2.941° of the 2θ range. In the primary optics, a 0.5-mm divergent slit and 2.5° Soller slits were mounted, while the detector optics with the detector window fully opened were limited to 2.5° Soller slits. The measurement range was 2θ = 20°–100° while the samples were spun. For annealed samples, an additional DP was collected in the 2θ = 25°–50° range where weak peaks were spotted, in order to ease the analysis of the phase(s) in this scattering region. The Fityk program [56] was used to evaluate the DPs. Combined pairs of Voigt functions, accounting for $K_\alpha 1$ and $K_\alpha 2$, were used in the analysis of diffraction peaks' profiles.

For in-operando powder X-ray diffraction (PXRD) experiments, the stability of the crystal structure of M2.S1 as a function of temperature was studied in the Anton Paar HTK 1200 N chamber. The chamber has an environmental heating mantel on all sides mounted on the alumina table with an attached resistant temperature sensor, and is equipped with a water-cooled housing, fused alumina insulation. DPs were collected in the 2θ = 25°–55° range at T = 25°C, 200°C, 400°C, 600°C, 800°C, 1000°C, and 1200°C. However, the study does not include temperatures above 800°C as the material lost its single-phase configuration. Since Si 4 0 0 reflection originating from the support was not accessible, the 2θ range was limited. Though the detector window was limited by the 5.3-mm antiscatter slit to avoid scattering artifacts, there was a primary slit with a variable width on the specimen's surface in order to form a 3-mm footprint of the beam.

The mechanical properties of the samples were determined by nanoindentation. Measurements were obtained at room temperature using a Berkovich-shaped diamond indenter. Indentations were performed in the load partial unload (LPU) mode. Thirteen different forces, ranging from 0.2 mN to 5 mN, were used to produce the LPU tests. Each indentation was repeated 15 times. The distance between the indents was 20 μm, which avoided the interference of the indents and indenting in an already deformed region. Using the Oliver and Pharr model [57], the nanohardness and reduced Young's modulus ($E_r$) were calculated from the nanoindentation load–displacement curves.

3. Results and discussion

3.1 EDS results

The chemical composition of samples was studied by EDS analysis, and the results are presented in Table 4. The findings showed relative changes in the atomic concentration as a function of the modulating frequency. Using the changes in the $f_{mod}$ value, the chemical composition of sputtered coatings can be varied, thus permitting optimization of the chemical composition of material without manufacturing new targets for each required composition. The results showed that the concentration of aluminum and nickel increase in response to high $f_{mod}$, while tungsten, titanium, and niobium remain constant (Figure 2.) The results of previous studies [31] showed, that the changes in modulation frequency in metallic coatings synthesis lead to significant changes in morphology and the growth rate of individual materials. These changes directly alter the chemical composition, leading to the entropy segmentation of sputtered coatings. HEA coatings with low $f_{mod}$ (M2.S1 and S3) exhibit a higher $\Delta S_{mix}$ (1.565 ± 0.007), compared to those synthesized with a higher $f_{mod}$ value (M2.S2 and S4) (1.51 ± 0.008). Additionally, the same grouping behavior can be seen with VEC and δ. Low $f_{mod}$ coatings are characterized by VEC = 5.32 ± 0.06 and δ = 5.37 ± 0.16, while high $f_{mod}$ coatings are characterized by VEC = 4.93 ± 0.02 and δ = 4.68 ± 0.21 (Figure 3. and Figure 4.). Minimal variation within groups in the PMS processes hints at the low power dependence of $\Delta S_{mix}$ and relative proportions of the elements. The $\Delta H_{mix}$ of samples cannot be grouped in terms of their sputtering modulation frequency. The only disparity seen in $\Delta \chi$ is dependent on whether a sample is of low or high entropy. However, within a group, the variation is nonexistent for low-entropy samples and minimal (30.5 ± 0.5) for high entropy samples.

| Sample | Al | W | Nb | Ni | Ti | $\Delta S_{mix}$ [-R] | $\Delta H_{mix}$ [kJ/mol] | VEC | δ [%] | $\Delta \chi$ |
|---|---|---|---|---|---|---|---|---|---|---|
| M1.S1 | 61 | 5 | 4 | 24 | 6 | 1.09 | -32.95 | 4.97 | 5.78 | 0.20 |
| M1.S2 | 64 | 5 | 4 | 22 | 5 | 1.05 | -31.11 | 4.82 | 5.59 | 0.20 |
| M2.S1 | 32 | 19 | 12 | 20 | 17 | 1.56 | -34.34 | 5.38 | 5.52 | 0.30 |
| M2.S2 | 39 | 20 | 12 | 13 | 16 | 1.50 | -32.96 | 4.91 | 4.74 | 0.31 |
| M2.S3 | 31 | 20 | 15 | 17 | 17 | 1.57 | -33.29 | 5.26 | 5.21 | 0.31 |
| M2.S4 | 38 | 18 | 13 | 14 | 17 | 1.52 | -34.05 | 4.95 | 4.88 | 0.30 |

Table 4. Chemical composition and corresponding values of $\Delta S_{mix}$ – mixing entropy, $\Delta H_{mix}$ – mixing enthalpy, valence electron concentration, δ [%] – atomic size difference and $\Delta \chi$ – electronegativity difference of synthesized samples.

In order to increase the precision of available tools, there is a need for further research on chemical composition calculations for PMS from mosaic targets.

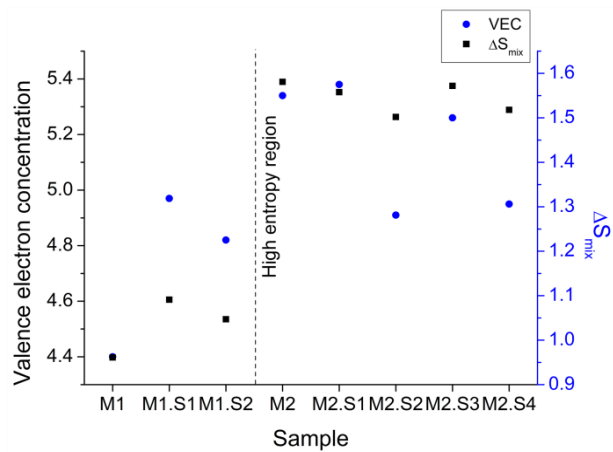

Fig. 3. Mixing entropy (ΔS$_{mix}$) and VEC plot of synthesized coatings.

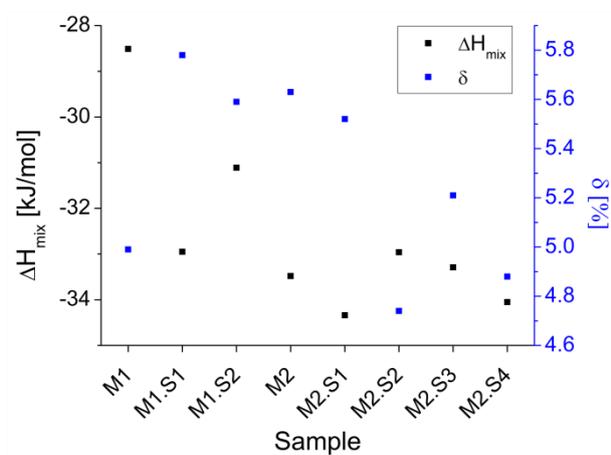

Fig. 4. Mixing enthalpy (ΔH$_{mix}$) and atomic size difference (δ) plot of synthesized coatings.

3.2 SEM imaging results

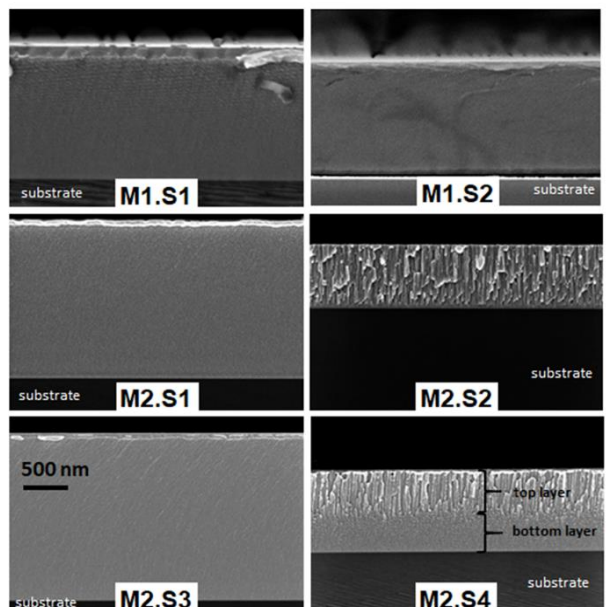

Fig. 5. Cross-sectional views of the as-deposited coatings.

| Sample name | Thickness [nm] | Growth rate [nm/min] |
|---|---|---|
| Mosaic 1 | | |
| M1.S1 | 1400 | 140 |
| M1.S2 | 1100 | 110 |
| Mosaic 2 | | |
| M2.S1 | 1700 | 170 |
| M2.S2 | 450 | 45 |
| M2.S3 | 2000 | 200 |
| M2.S4 | 750 | 75 |

Table 5. Coating thickness and growth rate.

It was observed in the SEM cross-sectional images (Figure 5.) that the as-deposited coatings had differences in their structures. All the samples exhibited smooth surfaces without any noteworthy features. The coatings obtained with frequency modulation of 10 Hz had a much smoother and denser morphology without any discernible structural features or defects. However, coatings obtained with frequency modulation of 1000 Hz also exhibited a dense compact columnar morphology that was formed during their growth. During the coating growth, the M2S2 coating showed such a hybrid structure, consisting of two sublayers—columnar (top) and dense (bottom)—after a period of time. The initial stages of coating growth resulted in a fine and dense coating, and under these conditions, columnar growth scenario was observed in the next step. The obtained structure with densely packed columns is typical for coatings synthesized with magnetron sputtering, similar to that of Thornton's T zone [58,59]. Such a structure was formed at higher ion bombardment during the deposition and thereby activated surface diffusion. The conditions that disrupt columnar growth (dense sublayer) are initially created during the synthesis under a higher power and higher modulation frequency. However, the thermal conditions change as the process proceeds due to the accumulation of heat in the material and the substrate. The change in the thermal conditions, along with the conditions of the plasma generation, caused the columnar growth. A similar effect was observed during the synthesis of Cu-N layers, when the power was increased [35]. It was observed that electrical conditions of the plasma generation caused differences in the thickness of the coatings. Based on previous studies, modulation frequency changes affect single time-life of plasma discharge and the total sputtering time. Using the PMS technique, modulation frequency allows for significant changes in morphology, phase composition, and the growth rate of synthesized thin films. The relationship in the Cu–N films between plasma composition and the modulation frequency was reported to have better control of the synthesis process, which allows for multiple possibilities in material design [36]. Similar relations have been noted in the synthesis of other materials, such as metals [31,32], oxides [60,61], and nitrides [33,36,62]. As observed in the present study, HEA coatings synthesized with a lower frequency (10 Hz) have more than twofold higher thickness.

3.3 XRD results

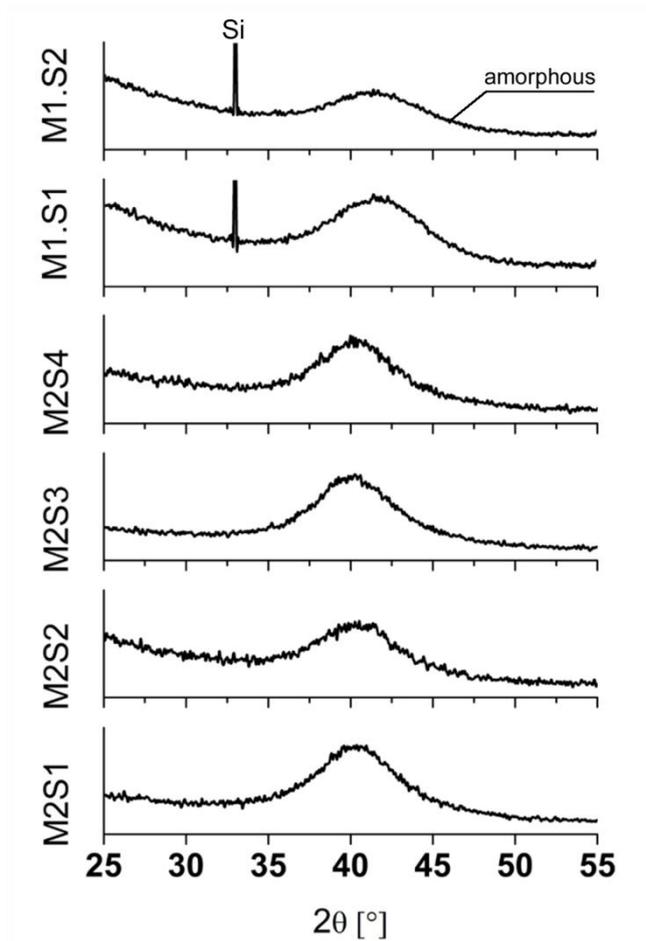

Fig. 6. XRD patterns of the AlWNbTiNi coatings as-deposited under various process parameters and from two kinds of mosaic targets.

Figure 6. presents our results of XRD measurements of as-deposited coatings synthesized from M1 and M2 targets. All the XRD patterns are identified as the broad halo pattern between 40° and 47°. XRD results show that all as-deposited samples are characterized by an amorphous structure.

| Sample | 2θ [°] | FWHM |
|---|---|---|
| Mosaic 1 | | |
| M1.S1 | 41.74 | 6.21 |
| M2.S2 | 41.75 | 5.83 |
| Mosaic 2 | | |
| M2.S1 | 40.32 | 5.93 |
| M2.S2 | 40.20 | 6.09 |
| M2.S3 | 40.37 | 6.32 |
| M2.S4 | 40.35 | 6.19 |

Table 6. Peak position (2θ), full width at half maximum (FWHM) of as-deposited coatings.

As observed previously [31–35], the pulsed processes on ambient temperature substrates create metastable states from plasma sputtered particles due to rapid heat loss, which effectively freezes atoms and particles on the substrate. The interaction of pulsed plasma with the surface creates an instantaneous and highly energetic thermal effect, which allows heat activation (for adhesion) and a limited short-range diffusion. Plasma-induced thermal interactions with the material synthesized using a high modulation frequency result in increased accumulation of heat. This can be seen as columnarization in Figure 5. (M2.S2 and M2.S4).

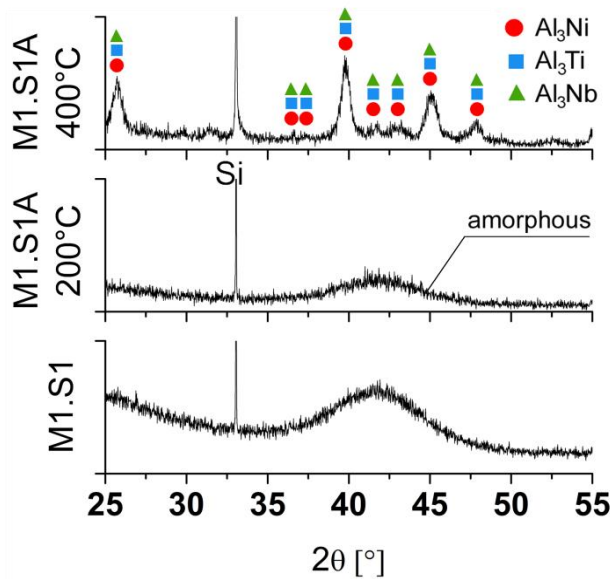

Fig. 7. XRD pattern of the thermal evolution of M1.S1.

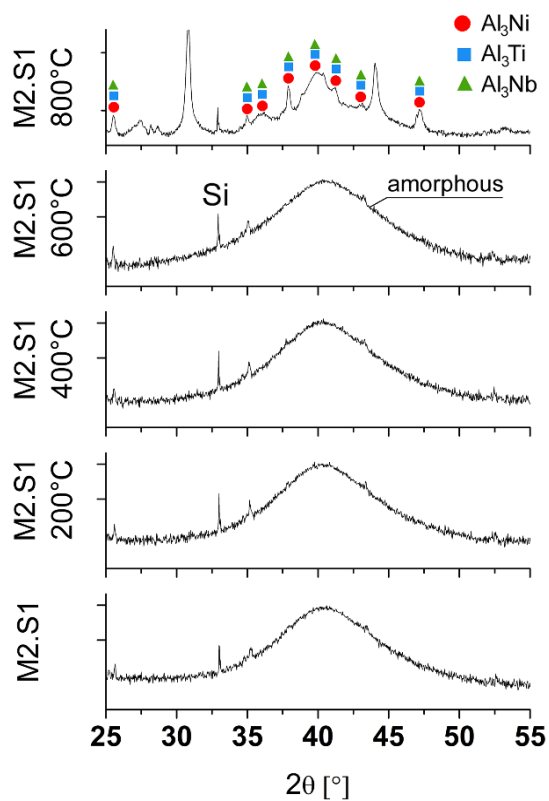

Fig. 8. XRD pattern of the thermal evolution of M2.S1.

| Sample name | M1.S1 | M1.S2 | M2.S1 |

| Temperature | 2θ [°] | FWHM | 2θ [°] | FWHM | 2θ [°] | FWHM |
|---|---|---|---|---|---|---|
| 200°C | 41.86 | 5.92 | 41.80 | 5.780 | 40.43 | 6.14 |
| 400°C | Decomposition into multiple phases | | | | 40.45 | 6.15 |
| 600°C | | | | | 40.57 | 6.53 |

Table 7. Peak position (2θ) and full width at half maximum (FWHM) of annealed coatings.

The results of thermal stability tests of representative metallic coatings, which were synthesized under identical pressure, power, and modulation frequency parameters of the M1 and M2 targets, are presented in Figure 7. and Figure 8. M1.Sx coatings obtained from Mosaic 1 (Figure 8.) retained their amorphous structure up to 400°C. At higher temperatures, the samples were subjected to relaxation, crystallization, and phase transitions due to further annealing.

In the M2.S1 coating (Figure 9.), the influence of high entropy—hindered diffusion—was evident, as the sample retained its amorphous configuration at significantly higher temperatures compared to the M1.Sx samples. During thermal evolution in the amorphous region, there were no differences in the XRD diagram between the M1.S1 and M1.S2 samples. Similarly, there was no variation seen in the M2.S1 coating between the heating stages up to 600°C, though it deteriorated into a multiphase structure at 800°C. Additionally, there was a similar transition observed in both cases, where the one-phase structure disappeared and peaks appeared at or in similar positions for 2θ, which could be attributed to several phases. The multiphase material likely contains trialuminide intermetallic phases such as $Al_3Ni$, $Al_3Ti$, $Al_3Nb$, or $Al_3(Ni, Ti, Nb)$. More peaks were found in diffraction spectra for coatings obtained from the M2 target. These untagged peaks (Figure 8.) are probably due to the high-temperature phases or due to surface oxidation during the XRD measurements at a heightened temperature.

For the Al-Ni alloy group, the $Al_3Ni$ eutectic phase is chemically and thermally stable, and it adds significant strength to aluminum and aluminum alloy through the well-known Orowan looping mechanism. Due to this, the alloys based on such structures are potential alternatives for traditional aluminum alloys for high-temperature applications [63]. During aging, the kinetics of $Al_3Ni$ precipitation and coarsening influence its hardness. Therefore, at elevated temperatures, materials acquire high strength from the precipitation of $Al_3Ni$ microfibers [64]. In this study, the formation of an $Al_3Ni$-like phase in the analyzed coatings strengthens the obtained structure of the material. The results of the mechanical properties reflect the above discussion.

3.4 Hardness results

Figure 9a. shows samples that are classical in their composition synthesized from the Mosaic 1 target, which have small differences in hardness both as as-deposited and annealed coatings. The values of M1.Sx average and M1.SxA average were 8.58 GPa and 8.31 GPa, respectively, and all the values lie within their respective error bands. More variation was observed in the M2.Sx family of samples. Since the error bars do not encompass all samples, a simple average cannot be calculated. Therefore, the samples were divided into two groups: the 10 Hz (M2.S1 and M2.S3) group and the 1000 Hz (M2.S2 and M2.S4) group. The 10 Hz group exhibited a noticeably higher hardness average, i.e., 10.97 GPa, compared to the 1000Hz group, i.e., 9.55 GPa. However, since the third grouping values lie within the error bar band, an average value from their hardness can sufficiently describe them, i.e., 14.03 GPa. The value of the third group was significantly higher than the 10 Hz pair (M2.S1 and M2.S3) and the 1000 Hz pair (M2.S2 and M2.S4), by 30% and 50%, respectively. The annealed M2.Sx samples exhibited an almost 70% increase in hardness compared to the average hardness of M1.Sx and M1.SxA.

The measured hardness values were well within the typical ranges of HEA coatings (5–15 GPa) [22,65–68].

As is known, various interdependent and competing phenomena such as phase transition, precipitation, and crystallization are responsible for overall microstructural evolution and hardening processes. As described by Haftlang and Kim [69], grain orientations, presence of different phases or precipitates in the microstructure can significantly affect the internal strain within a material and, subsequently, the strength – ductility relationship. The complex micro- and nanostructure of high entropy alloys allows for significant compositional fluctuations and phase reordering by undertaking various post-processing treatments (e.g. annealing). We can then attempt to control the mechanical properties of a material, including strength and ductility [69]. This is the result of the activation of various strengthening mechanisms, where some of them can occur simultaneously: solid-solution strengthening, composite effects, twinning, grain refinements, precipitation hardening [70,71]. Compared to conventional alloys, the diffusion rate in HEAs is lower due to the lattice distortion effect and the multi-principal component characteristics. Therefore, the diffusion needs to overcome the resistance caused by these two effects [72]. This phenomenon has a crucial effect on the nucleation and growth of precipitates in HEAs. Although the nucleation of precipitates in HEAs is easier, the growth as a rate-limiting step is slower when compared to the traditional alloys due to the very high interaction between elements. According to the literature, the precipitates formed in HEAs mainly include the $L1_2$ precipitate [73,74], $B_2$ precipitate [75], ρ precipitate [76], $D0_{22}$ precipitate [77], and l-phase precipitate [78]. The work by Haftlang and Kim [69] contains a more comprehensive review of these phenomena. Al is one of the main alloying elements used in the HEA structure, mostly used to form and stabilize the precipitates inside the HEAs' matrix. Furthermore, the presence of Al and/or Ti in the HEAs structure may enhance the mechanical properties through the formation of the B2 precipitate [79]. In case of our material the most likely explanation for this increase in hardness can be the solid-solution strengthening [80]. Another distinct possibility is precipitation hardening – described in section 3.3 XRD results discussion – as similarly significant increases in hardness were seen after annealing by Yurchenko et al. [81]. Additionally, it is also possible that nanometric and coherent precipitation of phases like B2 (e.g. TiAl) or $L1_2$ (e.g. $Ni_2TiAl$) is present, which would be impossible to detect with standard XRD measurements.

The uneven distribution between the 10 Hz (M2.S1 and S3) and 1000 Hz (M2.S2 and S4) pairs is due to their different structures. The 10 Hz-synthesized coatings had a uniform and fine structure, while the 1000 Hz pair exhibited a columnar or partially columnar structure (Figure 5.).

Figure 9c. shows the values of $H^3/E_r^2$ and $H/E_r$ representing the ratios of resistance to plastic deformation and wear resistance, respectively [82,83]. The M1.Sx samples exhibited a loss of close to 25% after annealing, while the M2.Sx samples were divided by $f_{mod}$ employed during sputtering. The 10 Hz samples exhibited a higher ratio than those sputtered under $f_{mod}$ = 1000 Hz. Annealing of the M2.Sx samples significantly increased their resistance indicators—M2.S1 and M2.S2—which showed an order of magnitude higher values. When comparing the as-deposited and annealed samples, a change in their behavior was observed. The as-deposited samples showed a higher value of H/E than $H^3/E^2$, while $H^3/E^2$ was higher after annealing.

Figure 9d. shows the plot of $1/E_r^2H$, which illustrates the resistance to crack damage [84]. The plot steadily decreases between the as-deposited and the annealed samples. There is a clear differentiation between the 10 Hz and 1000 Hz pairs of HEA coatings. However, there are opposing views on the use of $H/E_r$ and $H^3/E_r^2$ parameters to describe nonbrittle materials [85].

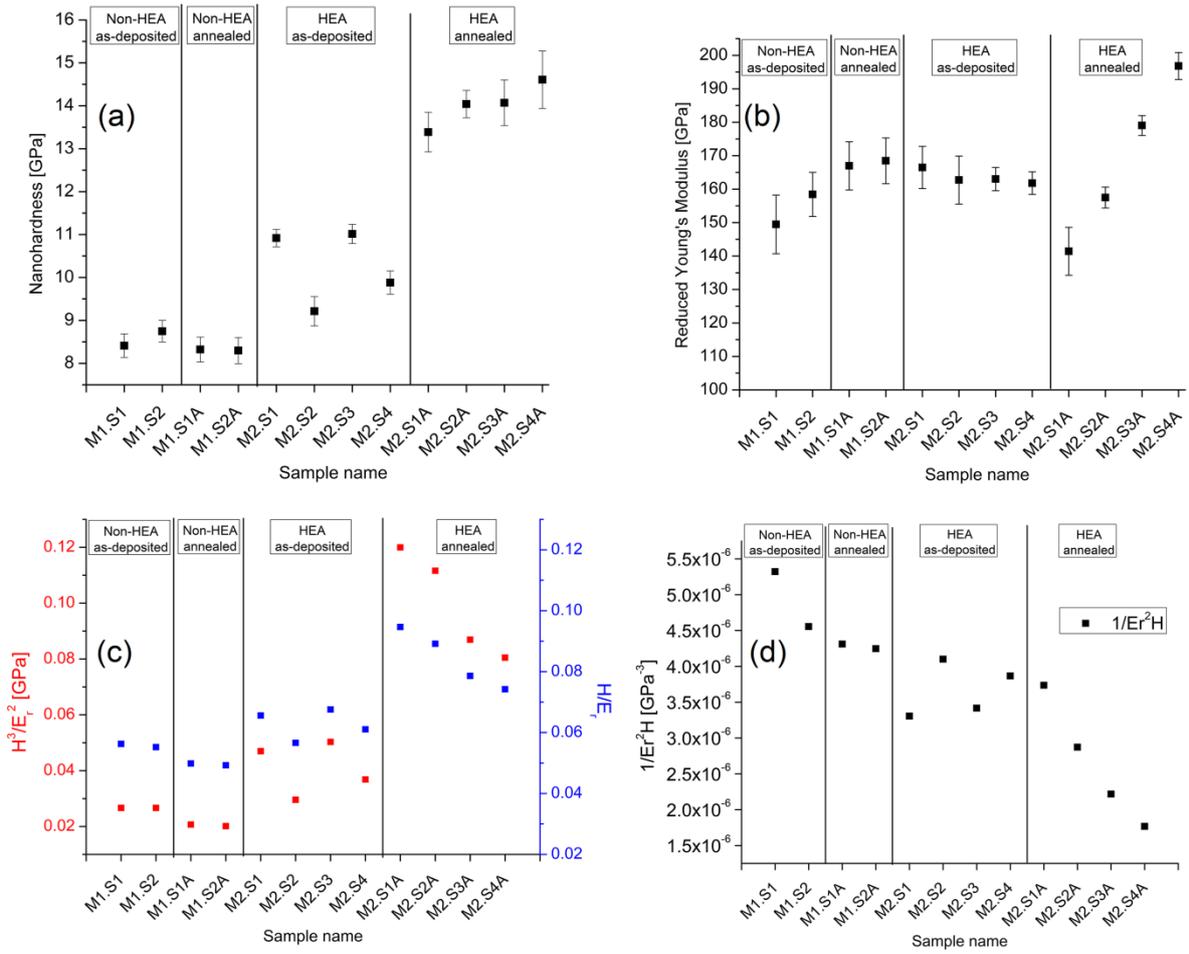

Fig. 9. Aggregated measurements for as-deposited and annealed M1.Sx and M2.Sx samples: (a) hardness (H), (b) reduced Young's Modulus ($E_r$), (c) $H^3/E_r^2$ and $H/E_r$, (d) $1/E_r^2 H$

Hardness measurements results (H, $E_r$) and derived calculations are presented in aggregate form in Table 6. for ease of use by other researchers.

| Sample | Hardness [GPa] | Hardness error | Reduced Young's modulus [GPa] | Reduced Young's modulus error | $H^3/E_r^2$ [GPa] | H/E | $1/E_r^2 H$ |
|---|---|---|---|---|---|---|---|
| M1.S1 | 8.41 | 0.27 | 149.47 | 8.76 | 2.66E-2 | 5.63E-2 | 5.32E-6 |
| M1.S2 | 8.75 | 0.25 | 158.43 | 6.60 | 2.67E-2 | 5.52E-2 | 4.55E-6 |
| M1.S1A | 8.32 | 0.29 | 166.96 | 7.20 | 2.07E-2 | 4.98E-2 | 4.31E-6 |
| M1.S2A | 8.30 | 0.305 | 168.48 | 6.84 | 2.01E-2 | 4.93E-2 | 4.25E-6 |
| M2.S1 | 10.92 | 0.20 | 166.46 | 6.30 | 4.67E-2 | 6.56E-2 | 3.31E-6 |
| M2.S2 | 9.21 | 0.34 | 162.72 | 7.17 | 2.95E-2 | 5.66E-2 | 4.10E-6 |
| M2.S3 | 11.01 | 0.22 | 163.02 | 3.46 | 5.03E-2 | 6.76E-2 | 3.41E-6 |
| M2.S4 | 9.88 | 0.27 | 161.80 | 3.38 | 3.68E-2 | 6.11E-2 | 3.87E-6 |
| M2.S1A | 13.39 | 0.46 | 141.41 | 7.17 | 12.00E-2 | 9.47E-2 | 3.74E-6 |
| M2.S2A | 14.04 | 0.32 | 157.47 | 3.12 | 11.16E-2 | 8.92E-2 | 2.87E-6 |
| M2.S3A | 14.07 | 0.53 | 179.02 | 2.95 | 8.69E-2 | 7.86E-2 | 2.22E-6 |
| M2.S4A | 14.61 | 0.67 | 196.82 | 4.02 | 8.05E-2 | 7.42E-2 | 1.78E-6 |

Table 8. Numerical values of H, $E_r$, H and $E_r$ measurement errors, and calculated values of $H^3/E_r^2$, H/E and $1/E_r^2 H$.

Analyzing Figure 9c., a pattern of $H^3/E_r^2$ versus H/E can be observed (Figure 10.). For as-deposited and annealed low entropy samples, the ratio of H/E is always higher than $H^3/E_r^2$. In the HEAs of the as-deposited samples, the pattern continues; however, after the annealing process, it is reversed with $H^3/E_r^2$ rising above H/E, though both values are higher than those of the as-deposited samples.

The $H^3/E_r^2$ and H/E values of the synthesized materials as well as their nonentropic variants are significantly higher than those reported in previous works [22,42,86–89]. Figure 10. shows a comparison of the best-case scenarios (highest values) of metallic samples obtained from previous studies, with M2.S3 (highest as-deposited) and M2.S1A (highest annealed) from this study. It can be understood that although there is no clear correlation between their values, when evaluating a material it is important to consider that the values vary greatly.

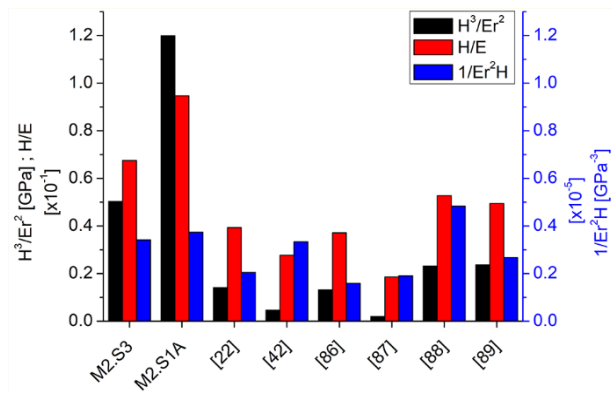

Fig. 10. Comparison between select $H^3/E_r^2$, $H/E_r$ and $1/E_r^2H$ from this work (M2.S3 and M2.S1A) and other works.

PMS is an efficient and effective technique, which creates alloyed coatings, both classical single-principal ones and new high-entropy ones with multiprincipal element compositions. Though the frequency modulation of plasma causes the formation of varied structures in reactive coatings such as Cu-N or Mo-N [33,35,37], it can also be used to design the structures of HEAs. Therefore, the properties such as hardness can be tailored to meet practical needs. Furthermore, $f_{mod}$ can likely be used within a fixed set of prefabricated targets in conjunction with mosaic targets to vary—to a limited extent—the chemical composition of sputtered coatings.

Lightweight HEAs present a wide range of possible applications. However, since intermetallic phases can emerge easily in these alloys, significant problems can arise. Our initial research and the proposed synthesis method showed that employing pulsed plasma magnetron sputtering in very low mixing enthalpy alloys is a promising venue of research. According to the obtained results, it is possible to synthesize single-phase amorphous structure in every studied case, while multiphase structure emerges after the annealing process. Additionally, control over obtained morphology and hardness of synthesized coatings can be achieved by using modulation frequency as a technological parameter.

4. Conclusion

In this study, multicomponent alloys of different chemical compositions were successfully synthesized by using the PMS method. Regardless of which mosaic target was used, all as-deposited coatings were characterized by a single-phase structure. However, their thermal stability was strictly dependent on their chemical composition. The single-phase composition was found to be stable up to 400°C for M1 alloys, and up to 800°C for M2 alloys.

The results of the study also showed that the process parameters used determine the morphology of the coatings and their properties. Low frequency modulations resulted in coatings with a fine structure and higher hardness compared to high frequency modulations which give rise to coatings with a columnar structure.

The mechanical properties of the studied HEA coatings, measured as a function of annealing temperature, revealed that annealing significantly increased the hardness of M2.S2 samples by more than 50% (from 9.2 GPa to 14.04 GPa).

Further study is needed on the change of $H^3/E_r^2$ with $H/E$ in annealed high-entropy samples, in order to understand whether this is only related to the presentation of data or if there is some underlying mechanism, such as precipitation strengthening, behind this transition.

The dependence of entropy on hardness in reference to the HEA coatings indicates that even small changes in entropy (<5%) can lead to significant changes in material properties—as observed between 10 Hz and 1000 Hz HEA pairs (M2.S1–S3 and M2.S2–4, respectively).

In conclusion, the presented results suggest that the use of PMS for the synthesis of HEAs is a promising venue of research into these chemically complex materials.

## 5. CRediT authorship contribution statement

G. W. Strzelecki – Conceptualization, Methodology, Formal analysis, Investigation, Data curation, Writing – original draft, Writing - revision, Visualization
K. Nowakowska-Langier – Supervision, Writing - original draft, Writing - revision, Analysis, Methodology, Resources
K. Mulewska – Investigation
M. Zieliński – Investigation, Writing - original draft
A. Kosińska – Investigation
S. Okrasa – Resources
M. Wilczopolska – Resources, Writing - original draft, Writing - revision
R. Chodun – Investigation
B. Wicher – Methodology
R. Mirowski – Resources
K. Zdunek – Supervision

## 6. Acknowledgements

M. Zieliński, A. Kosińska used equipment funded from the European Union Horizon 2020 Research and Innovation program under grant agreement no. 857470 and from the European Regional Development Fund via the Foundation for Polish Science International Research Agenda PLUS program grant no. MAB PLUS/2018/8.

## 7. Declaration of competing interest

The authors declare that they have no known competing financial interests or personal relationships that could have appeared to influence the work reported in this paper.